\documentclass[prc,aps,eqsecnum,twocolumn,floatfix,showpacs]{revtex4-2}
\usepackage{amsmath}
\usepackage{epsfig}
\usepackage{bm}
\usepackage{slashed}
\usepackage[active]{srcltx}
\def\tri{{{}^3{\rm H}}}

\def\het{{{}^3{\rm He}}}
\def\heq{{{}^4{\rm He}}}

\def\be{\begin{equation}}
\def\ee{\end{equation}}
\def\bea{\begin{eqnarray*}}
\def\eea{\end{eqnarray*}}

\def\bi{\begin{itemize}}
\def\ei{\end{itemize}}

\begin{document}

\title{Theoretical study of the $d(d,p)\tri$ and $d(d,n)\het$ processes at low energies}

\author{M. Viviani$^1$, L. Girlanda$^{2,3}$, A. Kievsky$^1$, D. Logoteta$^4$,
and L.E. Marcucci$^{1,4}$}

\affiliation{
$^1$ Istituto Nazionale di Fisica Nucleare, Sezione di Pisa, 
  Largo B. Pontecorvo 3, I-56127, Pisa, Italy \\
  $^2$ Department of Mathematics and Physics, University of Salento,
  Via Arnesano, I-73100 Lecce, Italy \\
$^3$ INFN-Lecce,  Via Arnesano, I-73100 Lecce, Italy \\  
$^4$ Department of Physics ``E. Fermi'', University of Pisa, 
Largo B. Pontecorvo 3, I-56127, Pisa, Italy }

\begin{abstract}
  We present a theoretical study of the processes $d(d,p)\tri$ and $d(d,n)\het$
  at energies of interest for energy production and for big-bang nucleosynthesis.
  We accurately solve the four body scattering problem using the {\it ab-initio} hyperspherical
  harmonic method, starting from nuclear Hamiltonians which include modern two- and
  three-nucleon interactions, derived in chiral effective field theory.
  We report results for the astrophysical factor, the quintet suppression factor,
  and various single and double polarized observables.
  An estimate of the ``theoretical uncertainty'' for all these
  quantities is provided  by varying the cutoff parameter used to regularize the chiral interactions
  at high momentum.
\end{abstract}


\maketitle

\section{Introduction}

The fusion reactions $d(d,p)\tri$ and $d(d,n)\het$
are critical processes for our understanding of Big-Bang nucleosynthesis (BBN)
and for new designs of fusion reactors.
In fact, the uncertainties in the prediction of the deuteron abundance [D/H] in BBN
models is currently dominated by the lack of precise knowledge of the astrophysical S-factor $S(E)$
of these processes~\cite{Yeh2021,Pisanti2021}. Therefore, accurate calculations of $S(E)$
could be very helpful in reducing the uncertainty of the [D/H] estimate.

Moreover, it has been speculated that the rate of $d(d,p)\tri$ and $d(d,n)\het$  would be
reduced preparing the initial deuterons with parallel spins (i.e. being in the ``quintet''
spin state)~\cite{Kulsrud1982,Gen10}. This suppression is referred as the quintet suppression.
The interest on this suppression is related to the construction of ``neutron lean reactors''with a  
$d+\het$ plasma, which would produce energy via the reaction $d+\het\rightarrow p+\heq$.
However, the neutrons from the process  $d+d\rightarrow n+\het$ would be always present.
Hence, the interest in the use of polarized fuel~\cite{Ciullo16} and in the quintet suppression.
Naively, the suppression of the $\vec d(\vec d,n)\het$   (and of $\vec d(\vec d,p)\tri$) rate is expected
when one assumes the capture to take place in S-wave. Then, the process would require a spin-flip to
produce either a $\tri$ or $\het$ nucleus, a process generally suppressed. 
However, this argument does not take into account the presence of
the deuteron D-state or the possible capture in P- and D-waves, whose importance has been
already established also at low energy~\cite{Gen10}.  The suppression factor of the reaction rate
when the two deuterons are in the total spin $S=2$ quintet state with respect to the unpolarized case
is referred as the quintet suppression factor (QSF). No experimental study of the QSF has
been reported so far. From the theoretical point of view, different predictions for the
QSF have been reported, all at variance between each other~\cite{Gen16}.
The most accurate calculations predict a mild rate reduction using a polarized beam
of laboratory energy above 50 keV, and even a rate increase at lower energy
(i.e. QSF $>1$)~\cite{Deltuva10}. Clearly, further studies are necessary
to better clarify this issue. 

Another advantage advocated for the use of polarized fuels in reactors, is related to the possibility of handling
the emission directions of reaction products, in particular the neutrons~\cite{Ciullo16}.
This could have an important impact on cost and safety of future
fusion reactors, having the possibility to design fusion chambers where less parts of the walls are
bombarded by neutrons~\cite{Gen10}. The PolFusion experiment is currently being designed to
study these processes using polarized deuterons for beam and target~\cite{polfusion11,polfusion20}.

The $d(d,n)\het$ reaction is also used as a source of neutrons, subsequently employed to produce innovative medical radioisotopes.
For example, the SORGENTINA-RF project~\cite{Sorge21} has been designed to use these neutrons
to produce ${}^{99}$Mo from the stable isotope ${}^{100}$Mo, via the reaction  ${}^{100}$Mo$(n,2n){}^{99}$Mo.
From ${}^{99}$Mo is then possible to produce ${}^{99m}$Tc, a radio-tracer used in single photon emission computed tomography.
Again, it is important to know accurately the corresponding $d(d,n)\het$ cross section in the
energy range more relevant for this application.

The general spin formalism for the scattering of two (identical) spin-one particles can be found in Ref.~\cite{Gen10}.
There are one unpolarized  cross section, one vector analyzing power, three tensor analyzing powers and 19
correlation coefficients.
For future reference, we consider the case of a deuteron beam of energy $T_d$ (in the lab. system), impinging on a deuteron
target at rest. The energy of interest for energy production is in the range $T_d=10-50$ keV, while for BBN
$T_d=100\div 400$ keV. For the production of ${}^{99}$Mo, a beam energy in the range $T_d=200\div300$ keV
is considered optimal. 

The total cross section (or equivalently, the astrophysical S-factor) has been studied with great detail,
in view of its importance for BBN and energy production. The most recent measurements are reported in 
Refs.~\cite{FRG85,Krauss87,Brown90,Greife95,Leonard06,Wang07,Tumino11,Li15,Li17}. However, 
as discussed earlier, the different sets of data show a fairly large scatter~\cite{Pisanti2021}.
The $d(d,n)\het$ astrophysical S-factor has been experimentally investigated also using
laser induced fusion in plasmas~\cite{Lattuada16}.  The unpolarized differential cross section measurements
reported in the literature are somewhat older
(and with a gap around $T_d\sim200$ keV)~\cite{Blair48,Volkov57,Ganeev57,Shulte72,Ying73,Brown90}.
Noticeably, there exist a few accurate measurements of vector and tensor analyzing observables below $T_d<100$ keV. 
In particular, very precise data for the tensor analyzing powers $A_{zz,0}$ and $A_{xx,0}-A_{yy,0}$
for both reactions $d(d,p)\tri$ and $d(d,n)\het$ have been reported~\cite{Fletcher94}.
Moreover, precise measurements of the $d(d,p)\tri$  $iT_{11}$, $T_{20}$, $T_{21}$, and $T_{22}$ observables
have been performed at the Tandem Accelerator Center at Tsukuba~\cite{Tagishi92}.
In all these cases, only the deuterons in the beam were polarized. The already cited PolFusion experiment
is planned to measure double-polarized observables, in particular $A_{z,z}$ and $A_{zz,zz}$~\cite{polfusion20}. 

The study of these processes demands accurate solution of the four nucleon scattering
problem, as S-, P-, and D-waves have been found to give important contributions, at low energy as well~\cite{Gen10}.
The importance of P- and D-waves may be understood by taking into account the large extension of the 
deuteron wave functions (still sizable at interparticle distances of $6$ fm).
Therefore, the two entrance particles will interact also at a relatively large impact parameter.

From the theoretical side, there are a few accurate calculations reported in literature, such as
those obtained from the solution of the Faddeev-Yakubovsky (FY)
equations~\cite{Deltuva10} and using the Correlated Gaussian method~\cite{Aoyama12}.
Other calculations can be found in Refs.~\cite{Fick83,Hofmann84,Uzu02}. 

In the present paper, we study these processes using the hyperspherical harmonics (HH)
expansion method~\cite{Mea20,Viviani20}.
The potentials considered in this study are the chiral nucleon-nucleon
(NN) interactions derived at next-to-next-to-next-to-leading order (N3LO) by Entem and
Machleidt~\cite{EM03,ME11}, with cutoff $\Lambda=500$ and $600$ MeV. We include in the Hamiltonian
also a chiral three-nucleon (3N) interaction, derived at next-to-next-to leading order (N2LO)
in Refs.~\cite{Eea02,N07}. The two free parameters in this
N2LO 3N potential, denoted usually as  $c_D$ and $c_E$, have been
fixed in order to reproduce the experimental values of the $A=3$ binding energies
and the Gamow-Teller matrix element (GTME) of the tritium $\beta$ decay~\cite{GP06,GQN09,Mea18,Mea18b}.
Such interactions will be labeled as N3LO500/N2LO500 and N3LO600/N2LO600.

We report here the results obtained for a selected set of observables and compare them with 
the available experimental data and other theoretical calculations. We also provide a 
preliminary estimate of the associated ``theoretical uncertainty'', calculated from the difference
of the results obtained with the two values of cutoff $\Lambda$. In future, we plan to perform
a better estimate of this uncertainty following the procedure of Ref.~\cite{Weso21}. However, we are confident that
the reported theoretical uncertainty be of the correct order of magnitude. This uncertainty takes
into account our incomplete knowledge of the nuclear dynamics. 

The paper is organized as follows. In Section~\ref{sec:theory} a 
brief description of the method is given, while in Section~\ref{sec:res} the results of the calculations
are reported and compared with a selected set of available experimental data.
The conclusions and the perspectives of this
approach will be given in Section~\ref{sec:conc}.

\section{Theoretical analysis}
\label{sec:theory}
In the following, we will denote with the index $\gamma$ a particular
clusterization $A+B$ of the four-nucleon system in the asymptotic region.
More specifically, $\gamma=1,2,3$ will correspond to the $p+\tri$, $n+\het$, and $d+d$ clusterization, respectively.
Please note that at the energies considered here, all these three asymptotic channels are open, while breakup 
channels are closed.
Let us consider a scattering state with total angular momentum quantum number
$JJ_z$, and parity $\pi$. The wave function $\Psi_{\gamma L S,JJ_z}$ describing a state with
incoming clusters $\gamma$ in a relative orbital angular momentum $L$ and channel spin $S$ [note that
$\pi\equiv(-)^L$] can be written as
\begin{equation}
    \Psi_{\gamma LS,JJ_z}=\Psi_{\gamma LS,JJ_z}^C+\Psi^A_{\gamma LS,JJ_z} \ ,
    \label{eq:psica}
\end{equation}
where the core part $\Psi^C_{\gamma LS,JJ_z}$ vanishes in the limit of large inter-cluster
separations, and hence describes the system where the particles
are close to each other and their mutual interactions are strong.
We compute $\Psi_{\gamma LS,JJ_z}^C$ by expanding it over the HH basis~\cite{Mea20,Viviani20}.
On the other hand, $\Psi^A_{\gamma LS,JJ_z}$ describes the wave function in 
the asymptotic regions, where the mutual interaction between the clusters is
negligible (except for the long-range Coulomb interaction). In the
asymptotic region therefore the wave functions $\Psi_{\gamma LS,JJ_z}$ reduces to
$\Psi^{A}_{\gamma LS,JJ_z}$, which must be the appropriate asymptotic
solution of the Schr\"odinger equation. The functions $\Psi^A_{\gamma  LS,JJ_z}$
depend on the T-matrix elements (TMEs) ${}^JT^{\gamma,\gamma'}_{LS,L'S'}$, which are the amplitudes
for the transition between the initial state $\gamma,L,S$ to the final state $\gamma',L',S'$
for the wave with the specified value of $J$. 
Clearly, we are interested in the terms ${}^JT^{\gamma=3,\gamma'=1}_{LS,L'S'}$ and ${}^JT^{\gamma=3,\gamma'=2}_{LS,L'S'}$.
Full detail of the procedure adopted to determine $\Psi_{\gamma LS,JJ_z}^C$ and the TMEs
is reported in Refs.~\cite{Mea20,Viviani20}.

\section{Results}
\label{sec:res}

\begin{figure*}[tb]
\centering
\includegraphics[scale=0.65,clip]{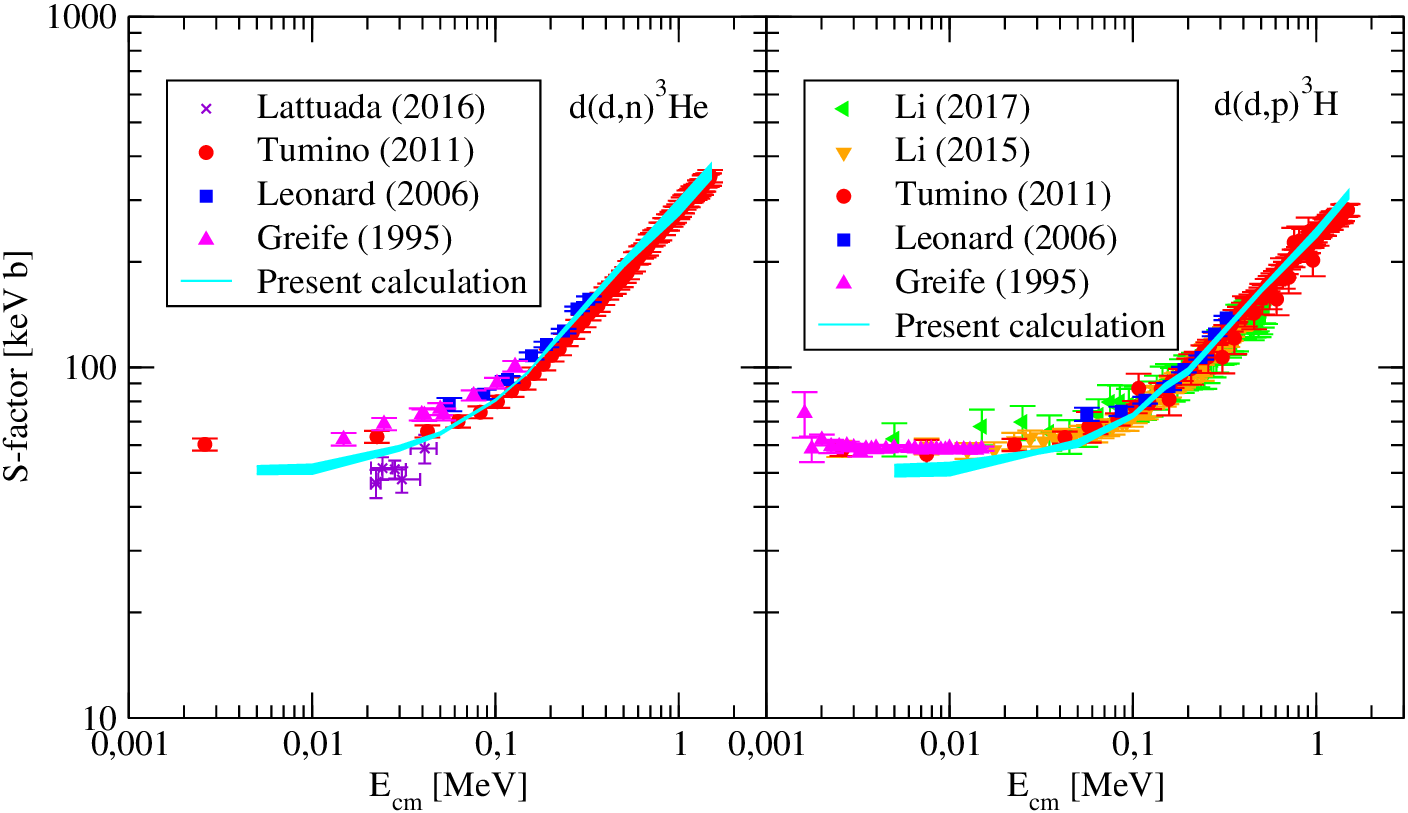}
\caption{(color online)
  The astrophysical S-factor for the processes $d(d,n)\het$ (left panel) and $d(d,p)\tri$ (right panel) calculated with the N3LO500/N2LO500
   and N3LO600/N2LO600 interactions. The width of the bands reflects the spread of theoretical
   results using $\Lambda=500$ or $600$ MeV cutoff values.
  See the main text for more details.
  The experimental values are from Refs.~\protect\cite{Greife95,Leonard06,Tumino11,Li15,Li17,Lattuada16}.}
  \label{fig:S}
\end{figure*}

First of all, let us consider the unpolarized total cross section, which is simply given by
\begin{equation}
  \sigma^{(\gamma')} = {1\over 6} {4\pi\over q_3^2} \sum_{J,LS,L'S'} (2J+1) |{}^JT^{(3,\gamma')}_{LS,L'S'}|^2\ ,
\end{equation}
where $q_3$ is the relative momentum between the two deuterons and
$\gamma'=1$ ($2$) for the $d(d,p)\tri$  [$d(d,n)\het$] reaction. We have calculated
it including all waves up to $L=4$. At $T_d<100$ keV, the dominant contributions comes from the
$L=0$ TMEs, ${}^0T^{(3,\gamma')}_{00,00}$ and ${}^2T^{(3,\gamma')}_{02,2S}$, with $S=0,1$
(the TME ${}^2T^{(3,\gamma')}_{02,21}$ gives the largest contribution).
However, there is also a sizable contribution from the TME
${}^1T^{(3,\gamma')}_{11,11}$, which, as the energy increases ($T_d>100$ keV)  becomes dominant.
Other $L=1$ TMEs contribute only marginally, while the $L\ge 2$ TMEs are much smaller and become sizable
only at $T_d\ge 1$ MeV. 

\begin{figure}[tb]
\centering
\includegraphics[scale=0.35,clip]{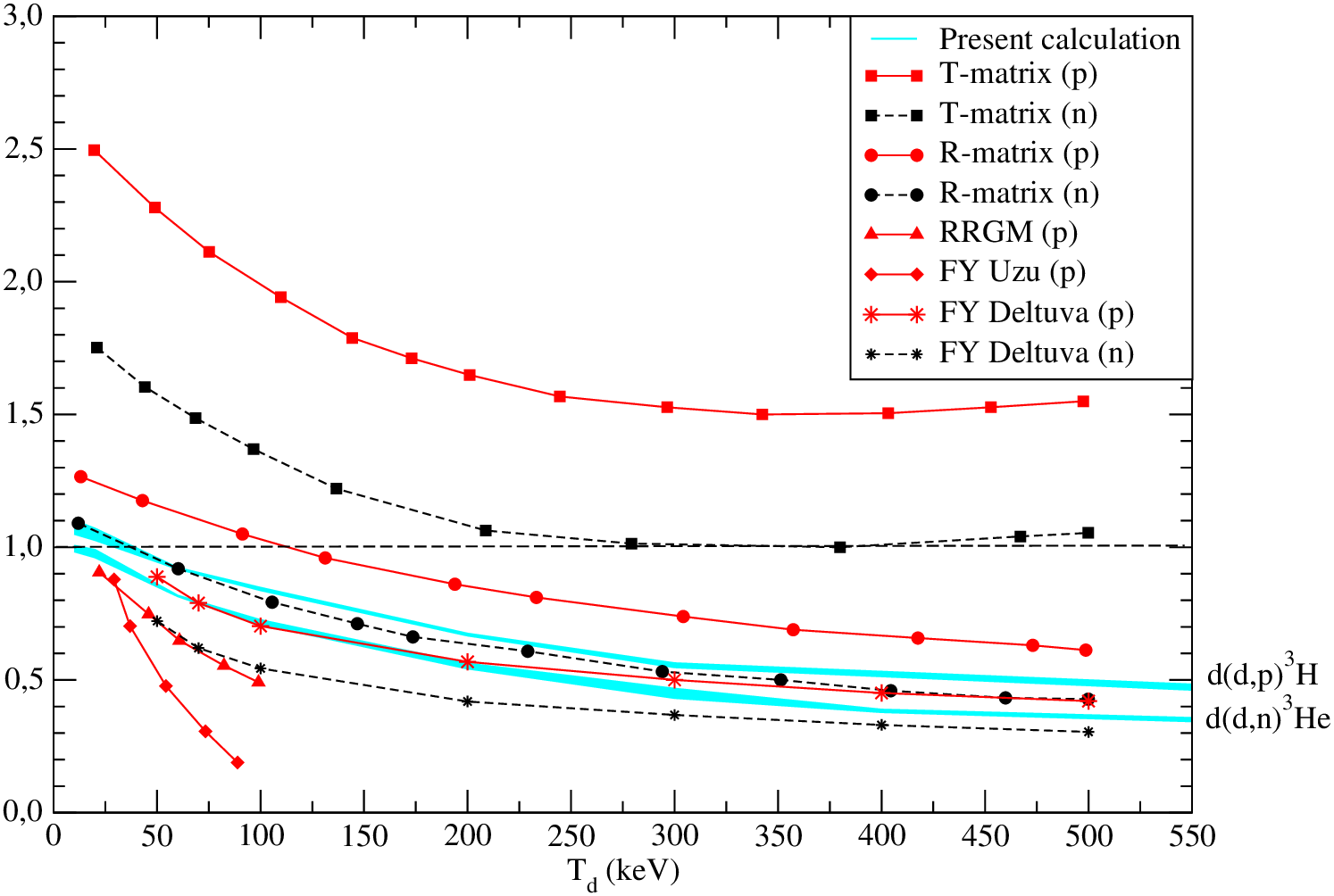}
\caption{(color online) The QSF for the processes $d(d,n)\het$  and
  $d(d,p)\tri$ shown as bands, in analogy of
  Fig.~\protect\ref{fig:S}. We report also the results obtained with
  other theoretical approaches: T-matrix~\protect\cite{Lemaitre93};
  R-matrix~\protect\cite{Fletcher94}; RRGM~\protect\cite{Fick83,Hofmann84};
  FY Uzu~\protect\cite{Uzu02}; FY Deltuva~\protect\cite{Deltuva10}. The red solid [black dashed] lines connecting the
  red [black] symbols denote the QSF calculated in the literature for the $d(d,p)\tri$ [$d(d,n)\het$] reaction. }
  \label{fig:qsf}
\end{figure}

From the total cross section, we have calculated the astrophysical S-factor,
defined as $S^{(\gamma)}(E_{cm})=E_{cm} \sigma^{(\gamma)} e^{2\pi\eta}$, where $E_{cm}=T_d/2\equiv q^2/2m$, $m$ being the nucleon mass
and $\eta=me^2/q$ the Sommerfeld parameter. 
The calculated S-factors $S^{(\gamma)}(E)$ for $\gamma=1,2$ are reported in Fig.~\ref{fig:S}, where they are compared with
recent experimental data~\cite{Leonard06,Tumino11,Lattuada16}.
The calculations have been performed using the N3LO500/N2LO500 and N3LO600/N2LO600
interactions and the results are shown as bands, their width reflecting
the spread of theoretical results using $\Lambda=500$ or $600$ MeV cutoff values.
As it can be seen from the figure, the calculations correctly reproduce the energy dependence of
the data. The astrophysical S-factor for $d(d,n)\het$ results to be larger than that of $d(d,p)\tri$ for $E_{cm}>0.1$ MeV. The calculations
are well in agreement with the data of Ref.~\cite{Tumino11}, while, the data of Ref.~\cite{Leonard06} are slightly underpredicted,
especially at low energy.

\begin{figure}[t]
\centering
\includegraphics[scale=0.45,clip]{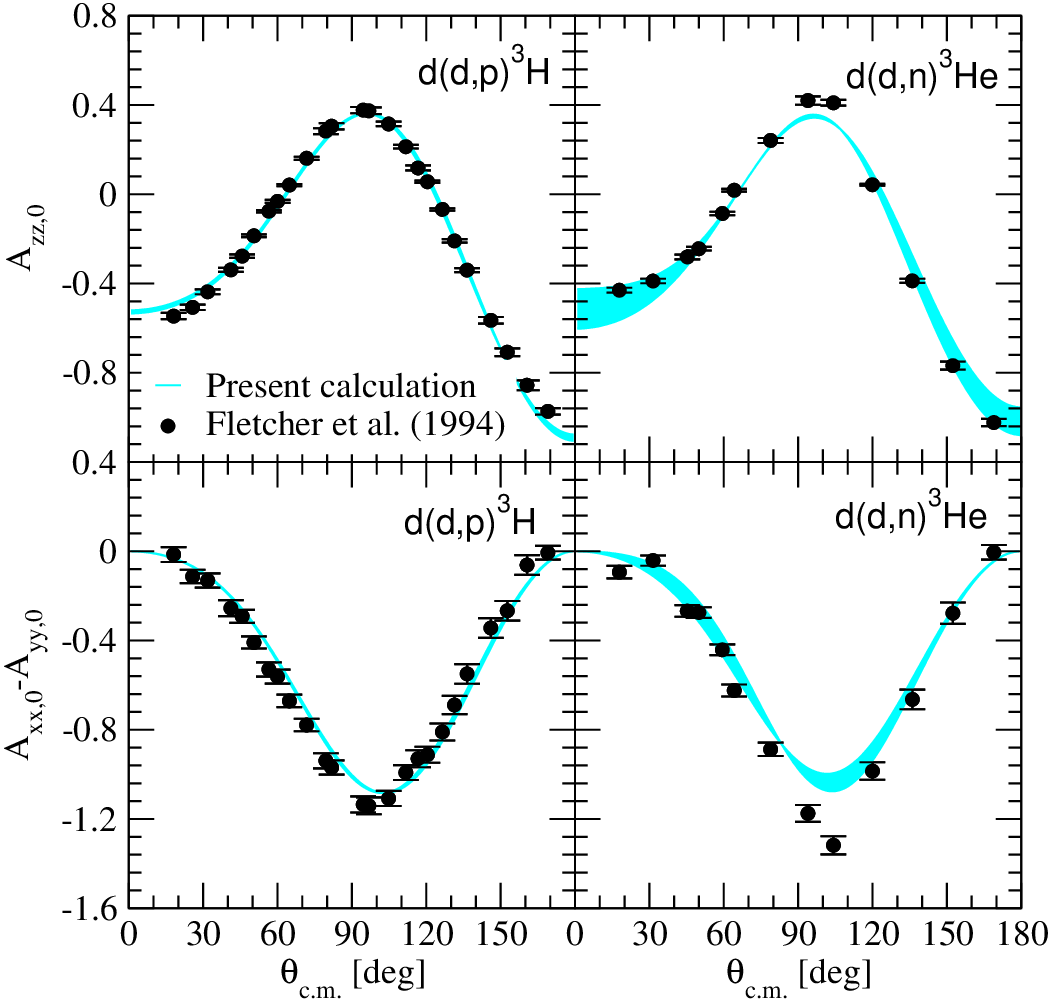}
\caption{(color online) The observables  $A_{zz,0}$ and $A_{xx,0}-A_{yy,0}$ for the  $\vec d(d,p)\tri$ and $\vec d(d,n)\het$ processes 
  at $T_d=21$ keV. The (cyan) bands show the results of the present calculations.
 The experimental  values are taken from Ref.~\protect\cite{Fletcher94}.}
  \label{fig:tap21}
\end{figure}

Next we consider the QSF. We compute $\sigma^{(\gamma)}_{11}$ as the total cross section for both deuterons
polarized along the beam direction. Then, QSF=$\sigma_{11}^{(\gamma)}/\sigma^{(\gamma)}$.
We report the calculated QSF in Fig.~\ref{fig:qsf}, together with other theoretical estimates
obtained using various methods~\cite{Lemaitre93,Fick83,Hofmann84,Uzu02,Leonard06,Deltuva10}. As it can be seen, our calculations
agree fairly well with the results of the FY calculation of Ref.~\cite{Deltuva10} and
with those obtained from the R-matrix analysis reported in Ref.~\cite{Leonard06}. Therefore,
the trend with energy appears to be well consolidated: the QSF is close to unity at small energies and then
slowly decreases. At $T_d=1$ MeV (not shown in the figure), it reaches a sort of plateau.
These findings are at variance, however, with what found by other analyses~\cite{Lemaitre93,Fick83,Hofmann84,Uzu02}.

\begin{figure}[t]
\centering
\includegraphics[scale=0.40,clip]{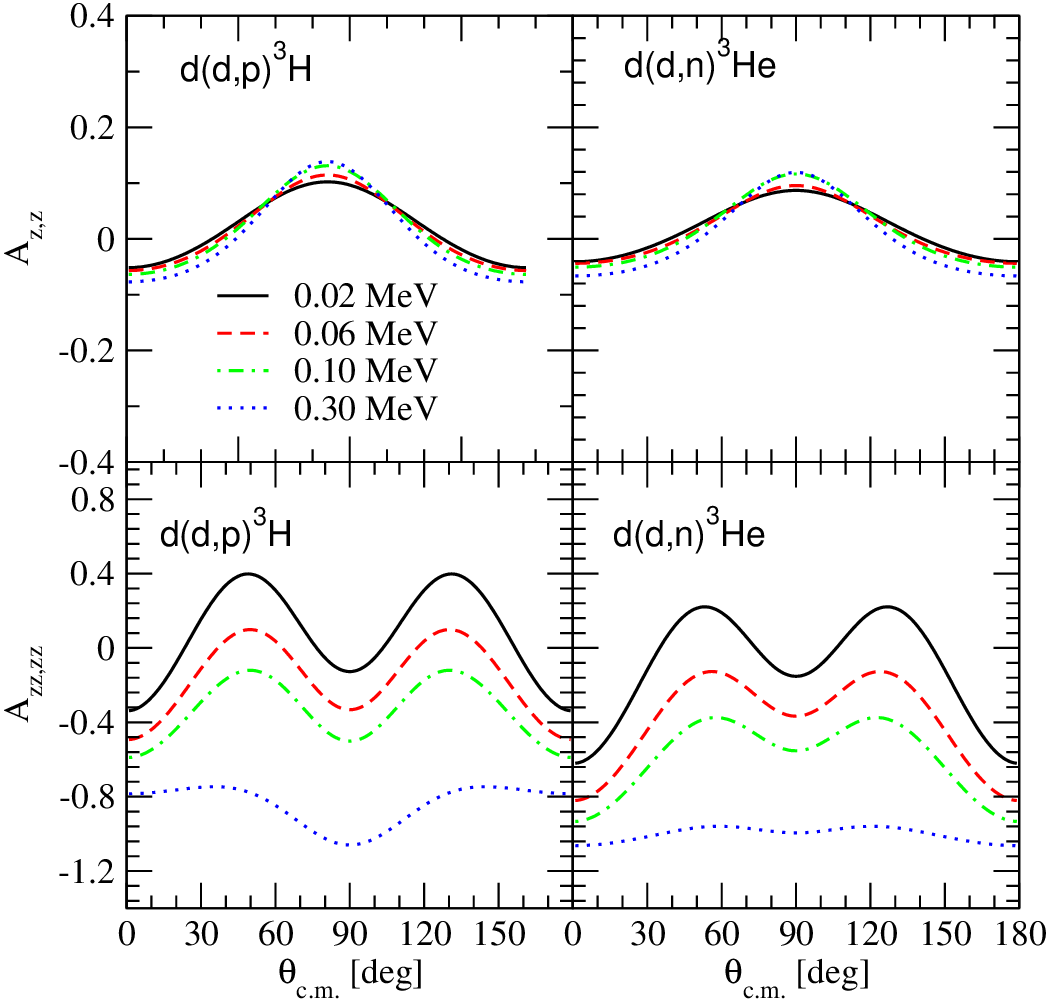}
\caption{(color online) The polarization observables $A_{z,z}$ and $A_{zz,zz}$ calculated for the
  $\vec d(\vec d,p)\tri$ and $\vec d(\vec d,n)\het$ processes at various laboratory energies.
The calculations have been performed for the N3LO500/N2LO500
  interaction. The associated theoretical error is of the order of 5\%}
  \label{fig:polfusion}
\end{figure}

The calculated unpolarized differential cross sections up to $T_d<1$ MeV, are generally in
good agreement with the experimental data~\cite{Blair48,Ganeev57,Volkov57,Brown90}.
More interesting is the comparison with the measured polarization observables below $T_d<100$ KeV.
For example, we report in Fig.~\ref{fig:tap21}, the comparison between our theoretical
results and the observables measured at $T_d=21$ keV in Ref.~\cite{Fletcher94}.
The results of our calculations are again shown as bands and they turn
out to be in good agreement with these experimental data.

We have performed other comparisons with the available
experimental data in this range of energies and a good agreement
between theory and measurements has always been found. 
We are therefore confident of the accuracy of the calculations and we
can make (sound) predictions for other observables. 
For example, in Fig.~\ref{fig:polfusion}, we show the prediction for the observables $A_{z,z}$ and $A_{zz,zz}$,
which will be studied in the near future by the experiment PolFusion~\cite{polfusion20}. The error estimated from the variation
of the cutoff in these cases is of the order of 5\%.

\section{Conclusions}
\label{sec:conc}
In this work, we have studied the $d(d,p)\tri$ and $d(d,n)\het$ processes at energies of
interest for BBN and for energy production in fusion reactors. The results of the
calculations have been presented as bands, being their width a preliminary estimate of the
theoretical uncertainty related to our incomplete knowledge of the nuclear dynamics.
In practice, the width of the bands reflects the difference between the theoretical results
obtained with the two values $\Lambda=500$ and $600$ MeV of the cutoff
parameter in the nuclear interaction. 
By taking into account the width of the bands, we can conclude that 
the theoretical results and the data well agree. We have also presented predictions for the QSF and
for some double-polarized observables, which will be the object
of a future campaign of measurements by the PolFusion experiment.
The  $d(d,p)\tri$ [$d(d,n)\het$] astrophysical S-factor at zero energy is estimated to be $S(0)=50.8\pm1.9$ keV b
($51.0\pm1.4$ keV b). The analysis of the consequences of these values
for the cosmological models is currently underway.

In future, we plan to perform a better estimate of the theoretical uncertainties, in particular, using
the new $\chi$EFT interactions derived up to next-to-next-to-next-to-next-to-leading order~\cite{MEN17}
and the procedure of Ref.~\cite{Weso21}. We plan also to study the changes in the fusion rates induced
by the presence of strong high-frequency electromagnetic fields, as there are suggestions
that the Coulomb barrier penetrability could increase significantly in certain configurations~\cite{laser1,laser2,laser3}.

{\it Acknowledgments} The Authors thank H. Karwoski and K. Fletcher
for providing them with their data.
The calculations were made possible by grants of computing time
from the Italian National Supercomputing Center CINECA and
from the National Energy Research Supercomputer Center (NERSC).
We also gratefully acknowledge the support of the INFN-Pisa computing
center. D.L. acknowledges the support of ACTA Srl, and in particular of
its CEO Dr. Eng. Davide Mazzini.
This article is supported by the Ministry of University and Research
(MUR) as part of the PON 2014-2020 “Research and Innovation"
resources – Green Action - DM MUR 1062/2021 of title
``Study of nuclear reactions of interest for the “green” energy production from nuclear
fusion''.

\end{document}